\documentclass{ws-ijmpcs}

\begin{document}

\markboth{Dimitrakoudis, Petropoulou and Mastichiadis}
{One-zone Hadronic Model}

%
\catchline{}{}{}{}{}
%
\title{THE TIME-DEPENDENT ONE-ZONE HADRONIC MODEL:\\ FIRST PRINCIPLES}

\author{STAVROS DIMITRAKOUDIS$^1$} 

\author{MARIA PETROPOULOU $^2$}

\author{APOSTOLOS MASTICHIADIS $^3$}

\address{Section of Astronomy, Astrophysics and Mechanics,
Department of Physics, University of Athens, Panepistimiopolis Zografos\\
Athens, Greece
\\
$^1$ st.seleukos@gmail.com\\
$^2$ maroulaaki@gmail.com\\
$^3$ amastich@phys.uoa.gr}

\maketitle

\begin{history}
\received{Day Month Year}
\revised{Day Month Year}
\end{history}

\begin{abstract}
We present some results on the radiative signatures of the one zone hadronic model.  
For this we have solved five spatially averaged, time-dependent coupled kinetic equations 
which describe the evolution of relativistic protons, electrons, photons, neutrons and neutrinos 
in a spherical volume containing a magnetic field. Protons are 
injected and lose energy by synchrotron, photopair and photopion production. 
We model photopair and photopion using the results of relevant MC codes, 
like the SOPHIA code in the case of photopion, which give accurate description 
for the injection of secondaries which then become source functions in their respective equations. 
This approach allows us to 
calculate the expected photon and neutrino spectra simultaneously
in addition to examining 
questions like the efficiency and the temporal behaviour of the hadronic models.

\end{abstract}

\ccode{PACS numbers: 11.25.Hf, 123.1K}

\section{Introduction}

Hadronic models for the high energy emission of
Active Galactic Nuclei and other astrophysical non-thermal objects postulate that protons are
accelerated by some process in the relativistic outflows connected with these objects.
The high energy photon emission comes then either from proton synchrotron radiation \cite{muckprot} or as
a  result of secondary particles
produced in photopion collisions -- for example, 
$\pi^0$ decay produces very high energy $\gamma$-rays, while it is possible 
that these high energy photons will be absorbed by soft photons,
initiating thus  intense electromagnetic cascades before escaping from the source \cite{mann}.

In the present paper we revisit the hadronic model 
by carefully modeling photopion production which is
one of its key processes. Furthermore we do not adopt the usual approach 
which is to assume an {\sl {ad hoc}} proton distribution in the source
but we go one step back and derive this as a solution of an equation 
which includes proton injection, losses and escape. 
This approach allows self consistency and in addition
it can adress some important points as is the efficiency
of the hadronic models and their expected time variability.

\section{Key assumptions of the model}

We adopt here  the standard picture of an one zone radiation
model. We assume that high energy protons are injected 
monoenergetically at some characteristic energy
$\gamma_0 m_pc^2$ in a spherical
source of radius $R$ which contains
a tangled magnetic field of strength $B$. 
We specify the rate of proton injection through the 
compactness $\ell_p$ which is related to the 
proton luminosity $L_p$ through the relation
$$\ell_p= {{L_p{\sigma_T}}\over{{4\pi R m_p c^3}}}$$
where $\sigma_T$ the Thomson cross section 
and we also introduce a proton escape timescale $t_{p,esc}$
which denotes the removal of protons from the source due to
physical escape. Finally, in order to keep the free
parameters as few as possible, we  assume that there is no injection
of primary electrons. 

Relativistic protons in the source lose
their energy by photopair, photopion and synchrotron radiation.
The stable and long-lived products of these interactions
include electron/positron pairs, photons, neutrinos and neutrons.
Electron/positron pairs radiate predominantly by synchrotron 
and inverse Compton radiation, so they become sources of photons.
High energy photons can be absorbed by photon-photon annihilation
producing electron-positron pairs. Low energy photons,
produced through e.g. electron synchrotron radiation,
become targets for protons, electrons and $\gamma-$rays.
Neutrons, not being confined by the magnetic fields in the source
can either escape or interact with the photons before decaying 
back to protons. Finally, neutrinos are the most uncomplicated
component as they will escape from the
source essentially with their production spectrum. 

It is clear from this description that the hadronic model
can be a highly non-linear system in the sense that its key components
(i.e. protons, electrons and photons)
are strongly coupled. In order to study it, one needs to use the kinetic
equation approach where 
the evolution of protons, electrons, photons, neutrons and neutrinos
are described through time-dependent
partial integro-differential equations. Each species is coupled with
the others through suitably modeled, energy conserving 
reaction rates representing the various important to the problem 
radiative processes. This approach has been used extensively
in leptonic models\cite{maskirk97}, however in hadronic models it has thus far
limited applications. One important reason for this was the 
difficulty in modeling the production rates of the secondaries in 
photopair and photopion interactions.

A step towards this direction was taken by Ref.~\refcite{MPK}
who have modeled photopair interactions by using the MC results of 
Ref.~\refcite{protjohn}; however
they used simple $\delta-$function approximations 
for the photopion reaction rates \cite{maskirk95}. In the present 
treatment we relax this 
assumption by modeling photopion
using the full results of the SOPHIA code \cite{SOPH}
-- details will appear in Dimitrakoudis et al. in preparation.

\subsection{Radiative signatures}

In addition to the above five parameters
(i.e. $R,~B,~\gamma_0,~\ell_p,~t_{p,esc}$)
one needs to specify
initial conditions for the five unknown distribution functions
to fully determine the system. Without loss of generality, we
can set them equal to zero for $t=0$.
Then we can integrate the equations forward in time.
Therefore, 
for $t>0$ protons
will start accumulating in the source. At the same time they will lose energy
by synchrotron, photopair and photopion, while a fraction
will physically escape at a rate $t^{-1}_{p,esc}$ from the source region.

\begin{figure}[h]
\begin{center}
$\begin{array}{c@{\hspace{0.01in}}c}
\multicolumn{1}{l}{\mbox{\bf (a)}} &    \multicolumn{1}{l}{\mbox{\bf (b)}} \\ [0.cm]
\epsfxsize=2.75in \epsffile{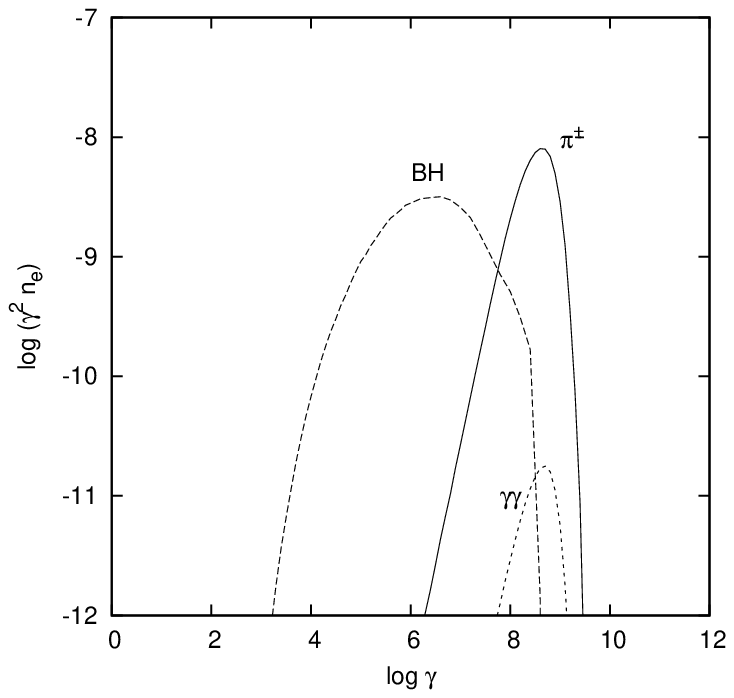} &    \epsfxsize=2.75in    \epsffile{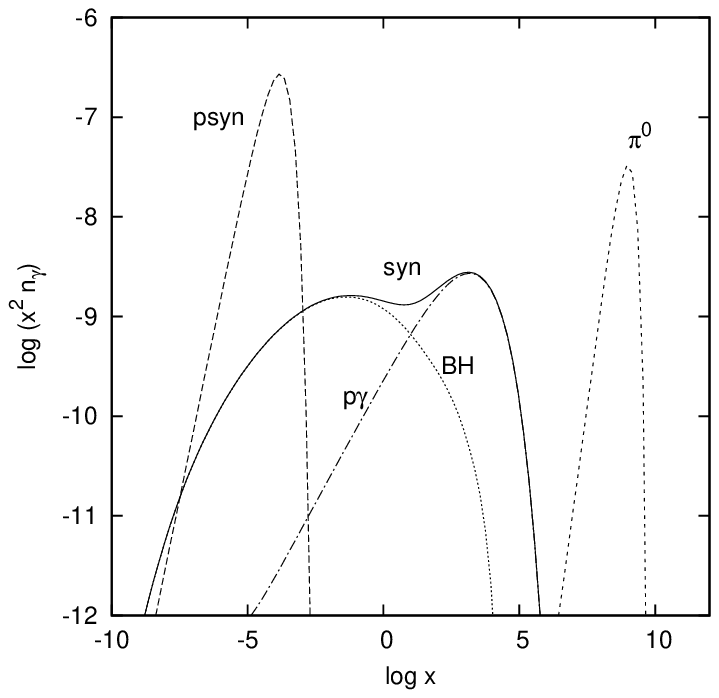} \\ [0.2cm]
\end{array}$
\end{center}
\caption{\small{(a) Production rate of secondary
electrons for
$R=3\times 10^{16}$~cm, $B=1~G$, $\ell_p=0.4$,
$t_{p,esc}=t_{cr}=R/c$ and $\gamma_0=2.5\times 10^6$.
`$\pi^\pm$' denotes the injection resulting from charged
pion decay, `BH' the photopair
(Bethe-Heitler) injection, while `$\gamma\gamma$' the injection
resulting from photon-photon annihilation
which is negligible for the particular
set of the initial parameters chosen.
(b) The corresponding steady state multiwavelength spectrum of photons resulting from the monoenergetic proton injection 
considered in (a). The `$p\gamma$' and `BH' curves are the synchrotron
spectra of the pairs injected by the photopion and photopair
processes respectively-- see (a). 
`psyn' is the proton synchrotron component while `$\pi^0$'
denotes the $\gamma-$rays resulting from the corresponding decay}}
\label{elecrate}
\end{figure}

Figure~1 shows a typical example where 
the parameters were chosen in such a way, so that the photons
radiated cause
minimal losses on the protons. Secondary electron/positron pairs
 are injected in the system mainly
through photopair and photopion.
(At higher compactness photon-photon pair production
becomes important but we can neglect this for the time being.)
These two processes are in direct competition with each other
and their relative importance depends on  $\gamma_0$
and on the soft photons which serve as targets.
Fig.~1a  plots the injection functions
of these two processes for the initial parameters given above.
We note that the two distributions have different characteristics.
The injection function of photopair electrons is broader and
has a peak at energies $\gamma_e\simeq\gamma_p$. The injection
function of photopion electrons, on the other hand, is flatter
and peaks at much higher energies, of the order of
$\gamma_e=\eta_{\pi e}\gamma_p$,
with $\eta_{\pi e}\simeq 150$.

In the case we are considering here, the photon spectrum
will show four distinctive features (see Fig.~1b). Two of them are conncected
to the synchrotron radiation of the injected pairs populations
discussed above while the other two features are connected to
proton synchrotron and $\pi^0-$decay respectively. 
Classifying them  in ascending
order with respect to frequency we have:

\begin{enumerate}

\item {{\sl Proton synchrotron radiation:} Since the proton distribution
function is a $\delta-$function at $\gamma_0$, the radiated photon
spectrum will have a peak at $\epsilon\simeq
{{m_e}\over{m_p}}b\gamma_0^2$.}

\item{{\sl Synchrotron radiation from photopair electrons:}
As stated above, the electron injection function resulting
from photopair interactions is rather broad with a peak at
$\gamma_e\simeq\gamma_p$. Synchrotron cooling of electrons
and consequent radiation results in a photon spectrum with peak
at $\epsilon\simeq b\gamma_0^2$.}

\item{{\sl Synchrotron radiation from photopion electrons:}
In complete analogy to the photopair, the peak of this
distribution will be at $\epsilon\simeq b(\eta_{\pi e}\gamma_0)^2$.}

\item{{\sl$\gamma$-rays from $\pi^0-$decay:} A monoenergetic proton
distribution produces a well defined peak at
$\epsilon\simeq\eta_{\pi\gamma}\gamma_0$, with $\eta_{\pi\gamma}\simeq 350$.}
\end{enumerate}

\subsection{Increasing the injected proton compactness: From linear
to non-linear proton cooling}

We turn next to investigate the effects that the injected proton
compactness has on the photon spectra. In the case of pure proton
injection as the one we are considering here, there are -- up to a degree,
profound analogies to the synchrotron - SSC relationship of a leptonic
system. There the electrons radiate synchrotron photons and
consequently upscatter them through
inverse compton scattering interactions.
As long as the magnetic energy density dominates the synchrotron
photon density, then the system can be considered to be in the linear
regime. This situation changes when the synchrotron photon density
dominates and the system becomes non-linear leading to the well-known
Compton catastrophe.

In the case of hadronic systems one can find also parameter values
that make the 
system operate in the linear regime. Such is the case shown in Fig.~1:
Protons radiate by
synchrotron and the radiated photons are used as targets
for photopair and photopion production. As can be seen from
Fig.~1b, the proton synchrotron luminosity dominates
which means that the cooling, however small, is regulated
by this process.
Therefore the question which becomes relevant
is what happens to the system if
the proton injection luminosity is increased
further while the magnetic field value is kept
constant. This essentially would mean that the photon
density of the system increases over the magnetic
one, and as a result the photopair
and photopion losses/injection increase more than
the proton synchrotron ones. This occurs because while
the synchrotron luminosity depends only on the proton
density, the photopair and photopion luminosities depend
on both the proton and photon density. Since the latter
depends on the former, we conclude that
the above processes depend quadratically on the proton density.

The above situation can be exemplified  in
Fig.~2 which depicts
the steady state multiwavelength spectra in the case where the injection
proton compactness is increased by a factor of ~3 over
its previous value, 
while all the other parameters are kept constant.
One notices that as the injection
compactness increases,
the synchrotron component increases linearly while the photopair
and photopion increase quadratically. However for the last
adopted value of $\ell_p$ the system undergoes a transition and
the photon luminosity increases by a factor of ~$10^4$. This type
of abrupt transitions are caused by various feedbacks 
\cite{sternsven,kirkmasti92,sternetal,petroumasti11}
which operate in hadronic systems
and are an indication that the protons become supercritical
inside the source. As it was shown in the references given above
in that case electrons and photons increase in an autoregulatory manner
causing the protons to lose energy by photopair and photopion
and forcing them to move back to the subcritical regime.
The system then can either reach a steady state, as in the
example shown above, or show limit cycles\cite{MPK}.

\begin{figure}
\resizebox{\hsize}{!}{\includegraphics[height=1cm,width=.5cm,angle=270]{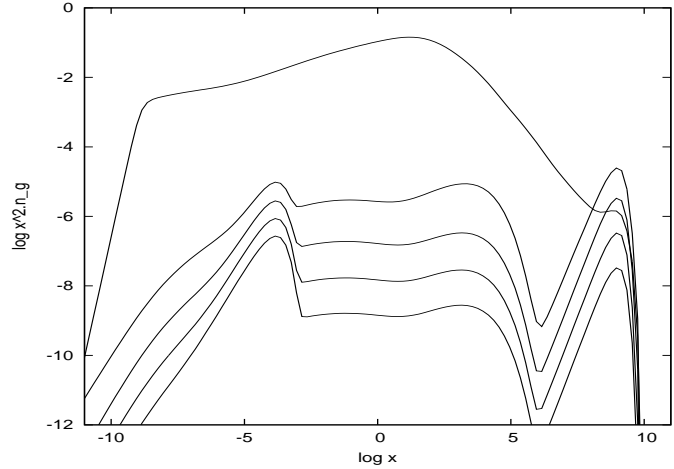}}
\caption{\small
{Steady state MW photon spectra for $\delta-$function
proton injection at energy
$\gamma_0=2.5\times 10^6$ and different injection compactnesses
$\ell_p=$0.4,~1.3,~4,~13 and 40 (bottom to top).
The other parameters are
$R=3\times 10^{16}$~cm, $B=1~G$, $t_{p,esc}=t_{cr}$. }}
\label{deltalp}
\end{figure}

Fig.~3 shows the locus separating the subcritical and supercritical 
proton regimes for various monoenergetic proton energies $\gamma_0$.
The concept is that once the proton energy density crosses this
boundary, protons become supercritical and therefore cannot
be stable for this value. It is interesting to note that, as preliminary 
calculations show, depending on the value of $\gamma_0$, 
different types of
feedback operate on the hadronic system. This explains
the shape of the locus. We will deal fully with the implications
of the hadronic non-linearities in a forthcoming publication. 

\begin{figure}
\begin{center}
\resizebox{3in}{!}{\includegraphics[height=4in,width=4in,angle=270]{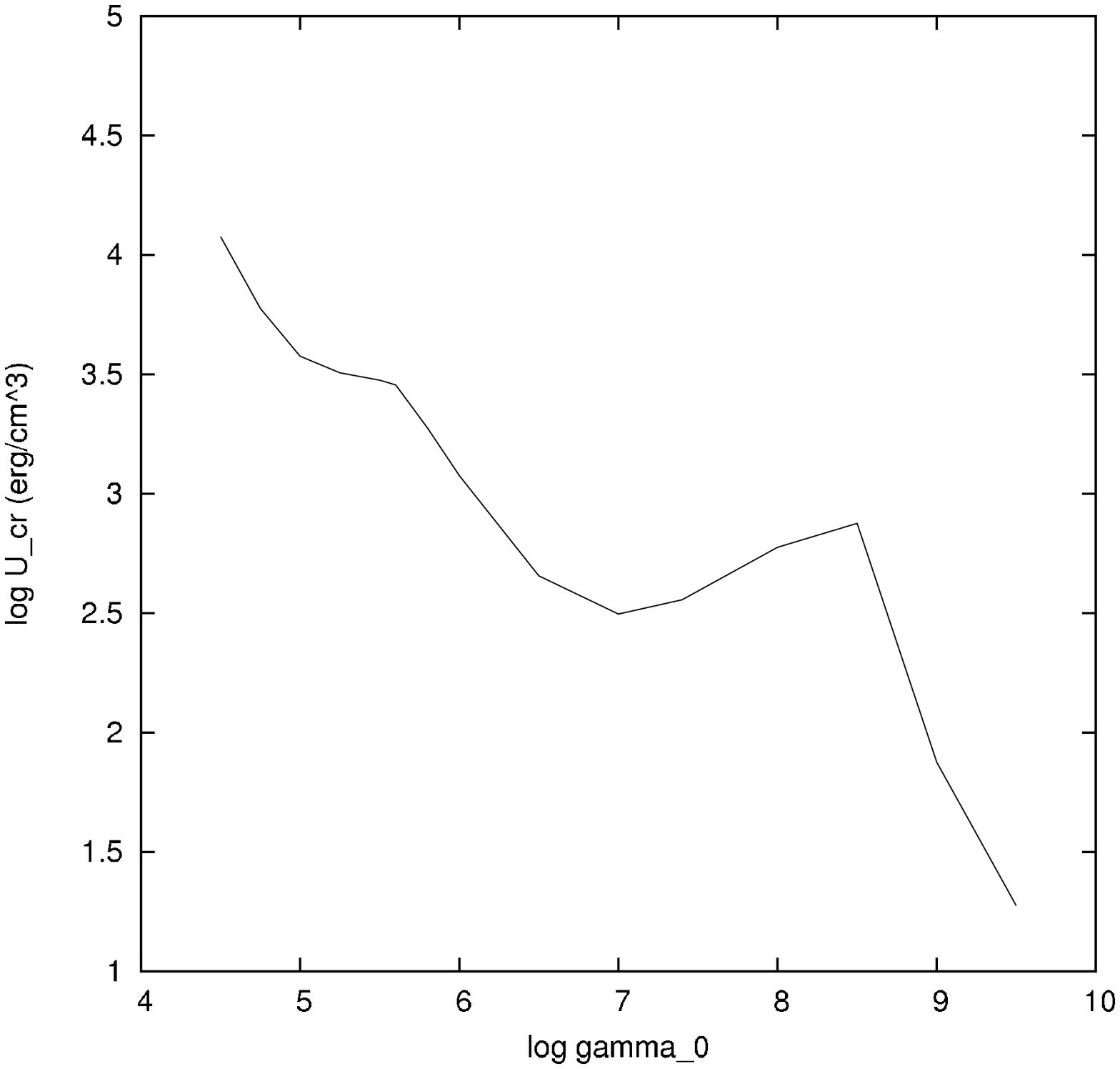}}
\end{center}
\caption{\small
{
Plot of the locus which separates the subcritical (lower part)  and supercritical
(upper part) proton regimes as a function of the proton energy $\gamma_0$
in the case where $R=3\times 10^{16}$ cm and $B=10~G$.
} }
\label{injectlp}
\end{figure}

{\sl{Acknowledgments}}: 
We would like to thank Drs R.J. Protheroe and A. Reimer
for making available the SOPHIA results to us.
This research has been co-financed by the European Union
(European Social Fund - ESF) and Greek national funds through
the Operational Program `Education and Lifelong Learning' of 
the National Strategic Reference Framework (NSRF) - Research
Funding Program: Heracleitus II. Investing in knowledge society
through the European Social Fund.




\begin{thebibliography}{0}    

\bibitem{muckprot} A. M\"ucke and R.J. Protheroe, {\it Astropart. Phys.}  {\bf 15} 121(2001).
\bibitem{mann}K. Mannheim, {\it Astron. \& Astrophys.} {\bf 269}, 67 (1993).
\bibitem{maskirk97} A. Mastichiadis, J.G. Kirk, {\it Astron. \& Astrophys.} {\bf 320}, 19, 1997.
\bibitem{MPK}A. Mastichiadis, R. Protheroe and J.G. Kirk,  {\it Astron. \& Astrophys.} {\bf 433} 765 (2005).
\bibitem{protjohn} R.J. Protheroe and P.A. Johnson {\it Astropar. Phys.} {\bf 4}, 253, (1996).
\bibitem{maskirk95} A.Mastichiadis and J.G. Kirk, {\it Astron. \& Astrophys.} {\bf 295}, 613 (1995).
\bibitem{SOPH} A. M\"ucke, et al., {\it Comp. Phys. Comm.} {\bf 124}, 290 (2000).
\bibitem{sternsven} B. Stern and R. Svensson,
Relativistic Hadrons in Cosmic Compact Objects, in {\it Lecture Notes in Physics} vol. 391, p.41 (Springer-Verlag Berlin Heidelberg New York 1991). 
\bibitem{kirkmasti92} J.G. Kirk and A. Mastichiadis, {\it Nature} {\bf 360}, 135 (1992).
\bibitem{sternetal} B. Stern, et al., {\it Mon. Not. R. Ast. Soc.} {\bf 272}, 291 (1995).
\bibitem{petroumasti11}
M. Petropoulou and  A. Mastichiadis, {\it  Astron. \& Astrophys.} {\bf 532}, 11 (2011).

\end{thebibliography}
\end{document}